\documentclass[aps,amssymb,twocolumn,showpacs]{revtex4}
\def\be{\begin{equation}}
\def\ee{\end{equation}}
\def\bq{\begin{eqnarray}}
\def\eq{\end{eqnarray}}
\def\bm{\begin{multicols}{2}}
\def\em{\end{multicols}}

\def\Hh{\mathcal{H}}
\def\Ee{\mathcal{E}}

\def\Oo{\mathcal{O}}

 1

\begin{document}

\title{Spin-fermion mappings for even Hamiltonian operators}

\author{Alberto Anfossi, and Arianna Montorsi}
\affiliation{Dipartimento di Fisica del Politecnico
di Torino, 10129 Torino, Italy}
\date{17 December 2004}

\begin{abstract}
\noindent We revisit the Jordan-Wigner transformation, showing
that --rather than a non-local isomorphism between different
fermionic and spin Hamiltonian operators-- it can be viewed in
terms of {\it local} identities relating different realizations of
projection operators. The construction works for arbitrary
dimension of the ambient lattice, as well as of the on-site vector
space, generalizing Jordan-Wigner's result. It provides direct
mapping of local quantum spin problems into local fermionic
problems (and viceversa), under the (rather physical) requirement
that the latter are described by Hamiltonian's which are even
products of fermionic operators. As an application, we specialize
to mappings between constrained-fermions models and spin 1 models
on chains, obtaining in particular some new integrable spin
Hamiltonian, and the corresponding ground state energies.
\end{abstract}  \maketitle

\paragraph{Introduction.}

Recently \cite{Ortiz,Ortiz_2} some more interest has been given to
the problem of mapping quantum lattice models of interacting spins
into fermionic lattice models, and viceversa, in the spirit of
unravel hidden structures (symmetries) of the problem by changing
its representation, and possibly identifying new integrable cases.
This is relevant for instance in the study of quantum phase
transitions, {\it i.e.} zero temperature changes of macroscopic
order induced by some interaction parameter.

The idea of spin-fermion mapping relies in fact on the old result
by Jordan and Wigner \cite{JOWI}, who first transformed quantum
spin $S=1/2$ operators, which commute at different lattice sites,
into operators obeying a Clifford algebra (fermions); the
transformation was used for mapping the one-dimensional XX model
into a spinless fermion model, the latter being exactly solvable.
The Jordan-Wigner transformation (JWT) was recently generalized in
\cite{ Ortiz} to cases of arbitrary spin $S$, which are naturally
mapped into multi-flavored fermions (for instance, electrons with
spin).

It is interesting to notice that the JWT is always a
\textit{non-local} transformation: in order to change the algebra
of the single particle operators at a given site (which is known
as transmutation of statistics\cite{Ortiz_2}), the transformation
in fact involves products of non trivial operators at each lattice
site. Nonetheless the JWT --being usually applied to Hamiltonian's
which are (sums of) even products of single particle operators--
turns out to transform \textit{local} Hamiltonian's into each
other; here, given a $D$-dimensional lattice $\Lambda$ with $L$
site, we define {\it local} any Hamiltonian ${\cal H}_{\bf j}$
acting on $n$ neighboring sites of ${\bf j}$ in $\Lambda$, such
that $\lim_{L\rightarrow\infty} {n\over L} =0$. JWT --when applied
to physically meaningful global Hamiltonian operators ${\mathcal
H} \doteq \sum_{\bf j} {\cal H}_{\bf j}$-- always maps a local
spin-spin term (say ${\cal H}^{(S)}_{\bf j}$) into a local
electron-electron term at the same lattice site ${\cal
H}^{(F)}_{\bf j}$. Such observation suggests that the mapping
induced by the non-local JWT could be fruitfully generated by just
a {\it local} transformation.

The spirit of the present paper is to unravel such local
transformation, and to give a systematic prescription which allows
to obtain directly at a local level both the results of
(generalized) JWT and new mappings between interacting spin and
electron Hamiltonian operators. This is useful for instance when
looking for integrable one-dimensional models: it is well known
that ${\mathcal H}$ does correspond to an integrable model
whenever the local matrix representing ${\cal H}_{\bf j}$ can be
expressed as a derivative of a $R$-matrix satisfying appropriate
equations (Yang-Baxter equations, see \cite{QISM} and references
therein). Hence integrability amounts to a local property:
understanding the local nature of the JWT should allow to generate
from a given local $R$-matrix both fermionic and spin integrable
models.

In order to achieve this goal, we first focus our attention on the
matrix representation of the 'isomorphic' operators related by the
JWT: the crucial and trivial observation is that they have in fact
identical matrix representation, meaning that they can be viewed
as different realizations --in terms of spin and fermionic
operators respectively-- of a unique formal operator. The latter
step will be achieved in the following through the combined use of
on site projection operators, and of the theory of matrix
representation for graded operators (see for instance
\cite{DOMO_NUCL} and references therein). After introducing the
reader to the method we then specialize to one dimensional case,
obtaining a single simple local equation relating spin $S$ models
to multi-flavored fermionic models, for arbitrary $S$. This is the
main result of our paper; the latter is shown to reproduces the
results of JWT for spin $1/2$ systems. We then explore in some
details the correspondence between constrained fermions models and
spin $S=1$ models; in particular, by this analysis we obtain some
new integrable spin 1 model, for which we explicitly derive the
ground state energy.

\paragraph{Matrix representation for even Hamiltonian operators.}

In order to obtain a matrix representation for a given Hamiltonian
operator $\Hh$, we have to specify the global vector space
$V^{(glob)}$ on which $\Hh$ acts. We begin by stating that this
space is a tensor product of $L$ copies of the local vector space
$V$ at each lattice site, \be V^{(glob)}=\underbrace{V \otimes
\ldots \otimes V}_{L\;times} \quad . \label{global_space}\ee Here
the order of the sites associated with the different copies of $V$
has to be fixed, meaning that the sites ${\bf j}$ on the $D$
dimensional lattice have to be put in one to one correspondence
with a scalar number $j$ ranging from $1$ to $L$. While such
correspondence is quite natural for $D=1$ (where ${\bf j}\equiv
j$), there are many possible different choices in case of
dimension greater than one (see for instance \cite{Ortiz_2}). We
shall not further enter here this argument, and in what follows we
simply assume that such correspondence has been set.\\Moreover,
assuming that $V$ has dimension $d$, we denote by
$|\alpha_j\rangle$ the $d$ state vectors which span $V$ at site
${j}$. These are defined through $d$ formal raising operators
$\{h^{(\alpha_j)}_j\}$'s which act on the local vacuum
$|0\rangle$, $|\alpha_{j}\rangle\doteq
h^{(\alpha_{j})}_{j}|0\rangle$. At this level, looking for
different realizations of $\Hh$ reduces to look for different
realizations of the $h^{(\alpha_{j})}_{j}$'s. In the following we
shall denote the different realizations by an index $X$ which can
assume both values $X=S$ for spin realizations, and $X=F$ for
fermionic ones; whereas, whenever we omit such index, we refer to
the abstract operators, {\it i.e.} $X$ can assume both values.
\\For instance, for spin realization, $V_S$ is spanned by the vacuum and the $d-1$
non vanishing powers of the raising operator $S_+$ of a spin ${\bf
S}$ operator with eigenvalue $S=(d-1)/2$; here the following usual
$su(2)$ commutation relations hold, \be
[S^{(j)}_+,S^{(j')}_-]=\delta_{j,j'} 2 S^{(j)}_z\quad , \quad
[S^{(j)}_z,S^{(j')}_{\pm}]=\pm \delta_{j,j'} S^{(j)}_\pm\quad .
\label{su(2)}\ee Whereas, in case of fermionic realization with
$d=2^{f}$ (with $f\in {\bf N}$ number of flavors of the fermions),
$V_F$ will be spanned by even and odd products of $f$ fermionic
creation operators $c^{\dagger}_{j,s}$, which satisfy the Clifford
algebra with anticommutation relations given by
\begin{eqnarray}
\{c^{\dagger}_{j,s}, c^{\dagger}_{j',s'}\} &=& 0 \quad ,\nonumber\\
\quad \{c^{\dagger}_{j,s}, c_{j',s'}\}     &=&
\delta_{j,j'}\delta_{s,s'} \; ; \quad s,s'=1,\dots,f \;
.\label{Clifford}
\end{eqnarray}
Due to the different (anti)-commutation relations for the
operators which realize $V$, the latter may ($V_F$) or may not
($V_S$) have an intrinsic graduation; in particular,
$V_F=V^{(0)}\bigoplus V^{(1)}$, where the odd (even) subspace
$V^{(1)}$ ($\,V^{(0)}$) is spanned by those vectors that are built
with an odd (even) number of creation operators. Similarly,
vectors and operators are said to have a parity $p=1$ ($p=0$).
\\With the above specifications we can write the basis vector of
the global vector space $V^{(glob)}$ as
\begin{equation}
| \alpha_1,\ldots \alpha_L \rangle \, \stackrel{def}{=}
\, h^{(\alpha_1)}_1 \ldots  h^{(\alpha_L)}_L | 0 \rangle \equiv
|\alpha_1\rangle\cdots |\alpha_L\rangle \label{state}
\end{equation}
where $| 0 \rangle $ is now the global vacuum, and the parity of
the above state vector is simply given by $\sum_j p(\alpha_j)$.

The Hamiltonian operator $\Hh$ is a global operator, defined on
the whole lattice. As specified in the introduction, we limit our
analysis to the case in which the Hamiltonian is a sum of local
operators $\Hh_j$, the latter acting on a vector space $V^{(n+1)}$
which is the tensor product of $n+1$ copies of $V$ on $n+1$
($n+1<L$) neighboring sites in the ordered state (\ref{state}).
Moreover, we require that $\Hh^{(F)}_{j}$ is a sum of even
products of fermionic operators, which implies that ${\Hh}_{j}$
has always parity $p=0$. The latter choice, which is quite
reasonable from the physical point of view, allows to limit the
problem of matrix representation of graded operators to that of
the matrix representation of just the local Hamiltonian
${\Hh}_{j}$. In fact, by using completeness and orthogonality
properties of the basis vectors (\ref{state}), we can rewrite
Hamiltonian operator $\Hh$ as \be {\Hh} = \sum_j
\sum_{\alpha_j,\dots,\alpha_{j+n}; \beta_j,\dots,\beta_{j+n}}
(H_{n+1})^{\alpha_j,\dots,\alpha_{j+n}}_{\beta_j,\dots,\beta_{j+n}}
\Oo_{\alpha_j,\dots,\alpha_{j+n}}^{\beta_j,\dots,\beta_{j+n}} \; ,
\label{ham}\ee where $H_{n+1}$ is a $d^{n+1}\times d^{n+1}$ matrix
representing the local Hamiltonian operator $\Hh_j$, and \be
\Oo_{\alpha_j,\dots,\alpha_{j+n}}^{\beta_j,\dots,\beta_{j+n}}\doteq
|\alpha_{j}\cdots\alpha_{j+n}\rangle\langle\beta_{j+n}\cdots\beta_{j}|
\label{op_prod}\ee are local projection operators acting on sites
from $j$ to $j+n$.\\
As expected, from (\ref{ham}) we have that the matrix
representation $H$ of the $d^L\times d^L$ global Hamiltonian
$\Hh$, which reads \be H = \sum_j{\bf I}\otimes \cdots \otimes
{\bf I} \otimes \underbrace{H_{(n+1)}}_{j\rightarrow j+n} \otimes
{\bf I} \otimes \cdots \otimes {\bf I} \quad , \label{H_matrix}
\ee is fully determined by the matrix representation of the local
problem, $H_{n+1}$. Hence, starting from a given $H_{n+1}$,
equations (\ref{ham}), (\ref{H_matrix}) establish that different
global isomorphic spin and fermionic Hamiltonian's are simply
obtained by looking to different realizations of local projection
operators (\ref{op_prod}). The latter contain the ultimate
significance of the JWT: by realizing them as fermionic or spin
projectors one obtains the corresponding local (and global) spin
and fermionic Hamiltonian. In the next paragraph, we shall
construct these operators (\ref{op_prod}) in terms of on-site
projector operators, which will then be implemented
explicitly in spin and fermionic realizations. \\
The above formulation of a quantum problem somehow reverse the
standard approach, which is to describe a physical problem through
an Hamiltonian operator, and successively to look for its matrix
representation in order to solve it. On the contrary here we start
from a matrix, which is the unique representation of a given {\sl
abstract} quantum problem --identified by the Hamiltonian
(\ref{ham}), and by the abstract projection operators
(\ref{op_prod})--, and look for its realization into different
operator languages, {\it i.e.} into different physical problems.
\\ Notice that such strategy would also hold for mapping to
operator languages other than spin or fermionic (for instance,
anyons or hard-core bosons) which can represent the (even)
Hamiltonian (\ref{ham}); the relevant point being the realization
of the local projectors (\ref{op_prod}) in the chosen language.
\\ Notice also that the local algebra obeyed by projectors
(\ref{op_prod}) is independent of the realization chosen. It reads
\be \Oo_{\alpha_j,\dots, \alpha_{j+n}}^{\beta_j,\dots,\beta_{j+n}}
\Oo_{\alpha_j',\dots,\alpha_{j+n}'}^{\beta_j',\dots,\beta_{j+n}'}
=
\delta_{\alpha_j',\beta_{j}}\dots\delta_{\alpha_{j+n}',\beta_{j+n}}
\Oo_{\alpha_j,\dots,\alpha_{j+n}}^{\beta_j',\dots,
\beta_{j+n}'}\label{dyn_alg} \, .\ee For $n$ and $d$ given, the
above relations close in a (sub)algebra (of) $u(d^{n+1})$ . Such
local algebra is characteristic of the abstract problem, and it is
realized through spin or fermionic operators when expressing the
even local operators $\Oo$ in terms of spin or fermionic
projectors.

\paragraph{Hamiltonian in terms of on-site projectors.}

Let us introduce the projection operators at site $j$,
$\Ee^\beta_\alpha\doteq |\alpha\rangle\langle\beta|$. The operator
in (\ref{op_prod}) can always be expressed as product of local
projection operators $\Ee_\alpha^\beta$. This is obtained by
combining together bra's and ket's at the same site, which, due to
the possible grading of the local vectors, requires some algebra.
In order to avoid cumbersome notation, we limit our analysis to
the case of a Hamiltonian which describes just two sites
interaction terms, the sites being at a distance $1$; such
assumption in fact limits the following analysis to the
one-dimensional case with nearest-neighbor interaction, though it
is easily generalizable to higher dimension, and to more general
local interactions. \\In this case the two-sites operator
(\ref{op_prod}) simply reads \be
\Oo_{\alpha_j,\alpha_{j+1}}^{\beta_j,\beta_{j+1}} =
(-)^{p(\beta_j) [p(\alpha_{j+1})+p(\beta_{j+1})]}
\Ee_{\alpha_j}^{\beta_j} \Ee_{\alpha_{j+1}}^{\beta_{j+1}} \quad ,
\label{op_proj}\ee where the sign in front of the on-site
projectors in fact can be negative only for a graded realization
of operator (\ref{op_prod}) (for instance, in the fermionic case).
By explicitly implementing (\ref{op_proj}) in the spin and
fermionic case, for {\it any dimension} $d$ of the on-site vector
space, one finally obtains the local mapping between the product
of on-site projection operators at $j$ and $j+1$ in the two cases,
which we may define as {\it local} (generalized) JWT. It reads \be
(\Ee_S)_{\alpha_j}^{\beta_j} (\Ee_S)_{\alpha_{j+1}}^{\beta_{j+1}}
\rightarrow (-)^{p(\beta_j)[p(\alpha_{j+1})+p(\beta_{j+1})]}
(\Ee_F)_{\alpha_j}^{\beta_j} (\Ee_F)_{\alpha_{j+1}}^{\beta_{j+1}}
 \label{JW_local} \ee

In practice, what is often convenient to do is:
\\($i$) start from
a known local problem in some operator language (spin or
fermionic);
\\($ii$) rewrite it in terms of on-site projection operator in the same
language; incidentally, this step gives the non-vanishing elements
of the local matrix $H_{2}$;
\\($iii$) use (\ref{JW_local}) to map products of on-site projectors in
the other language; and finally
\\(iv) look for the realization of
the projectors in the other language (fermionic or spin): this
will be the local Hamiltonian in the other language.

Here, as an example, we study the generic case in $d=2$.
\\We first construct the on-site projectors in terms
of spin $1/2$ and spinless fermions operators respectively. It is
useful to cast them into $2\times 2$ matrix $\Ee_X^{(2)}$ with
operator entries. In the spin case it reads \be \Ee^{(2)}_S =
\pmatrix{{1\over 2} - \sigma_z & \sigma_+ \cr \sigma_-&{1\over
2}+\sigma_z\cr}\quad , \label{proj_S}\ee where $\sigma_\alpha$
($\alpha=+,-,z$) are the Pauli spin $1/2$ operators, all with even
parity, and the on-site basis is spanned by the eigenvectors of
$\sigma_z$, $|-{1\over 2}\rangle$ and $|{1\over 2}\rangle$
respectively. In the fermionic case the projector matrix, in
terms of spinless fermion creation and annihilation operators
$c^\dagger$, $c$, reads \be \Ee^{(2)}_F = \pmatrix{1-n & c^\dagger
\cr c & n\cr} \quad , \label{proj_F}\ee with $n\doteq c^\dagger
c$; now the parity of diagonal entries is even, whereas that of
off diagonal entries is odd (in general, for graded vector spaces
$V^{(F)}$ the parity of projectors $\Ee_\alpha^\beta$ is
$p(\alpha)+p(\beta)$).\\The more general matrix representing an
even hermitian Hamiltonian with nearest neighbor interaction in
one dimension for $d=2$ is given by the $4\times 4$ matrix
$H_{2}^{(2)}$, \be H_{2}^{(2)}
=\pmatrix{h_{00}^{00}&0&0&h_{00}^{11}\cr
0&h_{01}^{01}&h_{01}^{10}&0\cr 0&h_{10}^{01}&h_{10}^{10}&0\cr
h_{11}^{00}&0&0&h_{11}^{11}\cr}\label{H2_matrix} \quad ,\ee where
the eight non-vanishing entries (arbitrary, except the constraints
$h_{01}^{10}=h_{10}^{01}$, and $h_{00}^{11}=h_{11}^{00}$ imposed
by the hermiticity requirement) have been written as
$h_{\beta_j\beta_{j+1}}^{\alpha_j\alpha_{j+1}}$ which makes it
easy to compare with (\ref{ham}). Now any local spin $=1/2$
Hamiltonian and its correspondent spinless fermion realization are
obtained from (\ref{ham}) and (\ref{op_proj}) by inserting a given matrix of the
form (\ref{H2_matrix}) and the appropriate projectors
(\ref{proj_S}) and (\ref{proj_F}) respectively.\\
For instance we may look at the fermionic realization of the
isotropic Heisenberg Hamiltonian for spin $1/2$ operators, which
is a sum of local two-sites operators of the form \be
\Hh_{H,j}^{(S)}= 2 \bar\sigma_j\cdot\bar\sigma_{j+1} \quad .
\label{Heisenberg} \ee It can be easily verified that it
corresponds to the choice $h_{00}^{00}=h_{01}^{10}= h_{11}^{11}=1$
and $h_{00}^{11}=h_{01}^{01}= h_{10}^{10}=0$ in (\ref{H2_matrix}),
in which case $H_{2}^{(2)}$ is just the standard two sites
permutation matrix; incidentally, this fact makes the model
integrable. With the above specifications, $\Hh_{H,j}^{(S)}$ can
now be written in terms of spin projectors, according to step (ii)
of our scheme, as
\begin{widetext}
\be \Hh_{H,j}^{(S)} = (\Ee_S)_{1_j}^{0_j}
(\Ee_S)_{0_{j+1}}^{1_{j+1}} + (\Ee_S)_{0_j}^{1_j}
(\Ee_S)_{1_{j+1}}^{0_{j+1}} + (\Ee_S)_{1_j}^{1_j}
(\Ee_S)_{1_{j+1}}^{1_{j+1}} + (\Ee_S)_{0_j}^{0_j}
(\Ee_S)_{0_{j+1}}^{0_{j+1}} \quad . \ee
\end{widetext}
By now implementing step (iii) and (iv) of the same scheme to map
bilinear products of spin into fermionic on-site projectors, and
realizing the latter through (\ref{proj_F}) in terms of fermionic
operators, we finally get the local Heisenberg Hamiltonian in its
fermionic realization, which reads \be \Hh_{H,j}^{(F)}=
c^\dagger_j c_{j+1}- c_j c^\dagger_{j+1} + 2 n_j
n_{j+1}-n_j-n_{j+1} \quad , \ee apart from constant terms. Of
course, such local identification reflects into an identification
of the global Hamiltonian's as well, the latter being in perfect
agreement with the results obtained by the non-local JWT.

\paragraph{Correspondence between spin $1$ models and extended $t-J$
models.}

Here we illustrate the correspondence obtained by the scheme
developed in the previous section in case of an on-site vector
space of dimension $d=3$. Such case is of particular interest
since the models obtained in both realizations are thoroughly
studied in the literature, exhibiting a rich structure of quantum
phase diagram; in the fermionic case, these are extended $t-J$
models with constrained fermions \cite{SCH}(in particular, $t-J$
model and infinite $U$ Hubbard model), whereas in the spin
realization they correspond to spin $1$ models.
\\In order to perform the local mapping, we have first to give
the on-site projectors in the two realizations. This requires to
specify which are the basis vectors of the on-site vector space
$|\alpha_R\rangle$ in the two cases, as well as how each of them
correspond to a different abstract $\alpha$.\\
For instance, in the previous example with $d=2$ we implicitly
assumed that the state $|-1/2\rangle$ in the spin realization was
implemented as the empty state in the spinless fermion
realization. Of course we could have chosen the opposite way,
associating the empty state with the state $|1/2\rangle$ in the
spin realization; if that was the case, we would have obtained a
particle-hole transform of the fermionic model corresponding to
the {\it same} spin model, meaning that such operation (which is a
mere redefinition of basis) does not change at all  the spectrum
of the model. In general, for $d>2$, one may obtain --associated
to different identifications of the corresponding basis vectors in
the two representations-- many different mappings of the same
(say) spin model, , whose number is given by $(d-n_1)! n_1!$,
$n_1$ being the number of odd-parity states ({\sl i.e.} the
dimension of $V^{(1)}$). \\In view of the above, we first proceed
to the separate construction of projection operators in the two
realizations, and only after we eventually specify a
correspondence between state vectors in the two realizations. In
the spin case, we choose $|1_S\rangle=|1\rangle$,
$|2_S\rangle=|0\rangle$, and $|3_S\rangle=|-1\rangle$, where the
index on the {\sl r.h.s.} refers to the eigenvalue of $S_z$. With
this choice the projector matrix in the spin $1$ realization reads
\be \Ee^{(3)}_S = {1\over 2}\pmatrix{S_z^2 +S_z & \sqrt{2}S_z S_+
& S_+^2 \cr  \sqrt{2} S_-S_z &2 (1-S_z^2) & -\sqrt{2} S_+ S_z\cr
S_-^2 & -\sqrt{2} S_z S_- & S_z^2- S_z} \quad .\ee For the
fermionic case, we may identify the basis of the on-site fermionic
Hilbert space with the three possible physical states
$|1_F\rangle=|\uparrow\rangle$, $|2_F\rangle=|0\rangle$,
$|3_F\rangle=|\downarrow\rangle$ (with  $p(2_F)=0$,
$p(1_F)=p(3_F)=1$), since two fermions on the same site are not
allowed (constrained fermions); with this choice, the projection
operators $\Ee_{F\, \alpha}^\beta$, which turn out to be a subset
of so called Hubbard projectors, can be cast again in the form of
a $3 \times 3 $ matrix $\Ee^{(3)}_F$ with operator entries;
explicitly, \be \Ee_F^{(3)} =\pmatrix{\tilde n_\uparrow & \tilde
c^\dagger_\uparrow  & \tilde c^\dagger_\uparrow \tilde
c_\downarrow \cr \tilde c_\uparrow & 1-\tilde n_\uparrow - \tilde
n_\downarrow & \tilde c\downarrow \cr \tilde c_\downarrow^\dagger
\tilde c_\uparrow &\tilde c^\dagger_\downarrow &\tilde
n_\downarrow} \quad , \label{ferm_proj}\ee where, as usual, we
have introduced the constraint of no double occupation through the
constrained fermion operator $\tilde c_\sigma\doteq
(1-n_{\bar\sigma}) c_\sigma$, with $\bar\sigma=-\sigma$; moreover
$\tilde n_\sigma\doteq \tilde c_\sigma^\dagger \tilde c_\sigma$.
The more general constrained fermions Hamiltonian with nearest
neighbor interaction is the $t-J-V$ Hamiltonian \cite{SCH}
\begin{widetext}
\be H_{tJV}^{(F)} = - t \sum_{j,\sigma}(\tilde
c_{j,\sigma}^\dagger \tilde c_{j+1,\sigma}+h.c.) + V \sum_j \tilde
n_j \tilde n_{j+1} + J\sum_j \vec{{\mathcal{S}}}_{j} \cdot
\vec{{\mathcal{S}}}_{j+1}\quad ,\label{tJV} \ee
\end{widetext}
where the standard on-site ($su(2)$) spin operator
$\vec{{\mathcal{S}}}_j$ has been introduced:
$\mathcal{S}_{+,j}=\tilde c_{\uparrow,j}^\dagger \tilde
c_{\downarrow,j}$, $\mathcal{S}_{-,j}= \mathcal{S}_{+,j}^\dagger$,
$\mathcal{S}_{z,j}={1\over 2} (\tilde n_{\uparrow,j}-\tilde
n_{\downarrow,j})$. In (\ref{tJV}) terms not conserving the total
number of electrons ${\mathcal{N}}\doteq\sum_j \tilde n_j$ and the
total spin operator $\vec{{\mathcal{S}}}\doteq \sum_j
\vec{{\mathcal{S}}}_j$ have been neglected.\\
Interestingly, $H_{tJV}^{(F)}$ reduces to the infinite $U$ Hubbard
model for $J=V=0$, and to the standard $t-J$ model for $V=-{J\over
4}$, both of which have been widely investigated in the
literature; in particular the exact analytical solution is known
in one-dimension both for the infinite $U$ Hubbard
model\cite{ISCA} and for the supersymmetric ({\it i.e.} $J=-2 t$)
$t-J$ model\cite{SCH,SUT}.\\
The spin $1$ realization of the $t-J-V$ Hamiltonian (\ref{tJV})
is now obtained by specifying which $\alpha_S$ corresponds to a
given $\alpha_F$; we choose $|\alpha\rangle_F\rightarrow
|\alpha\rangle_S$, in which case --up to conserved quantities--
the local JWT (\ref{op_proj}) gives
\begin{widetext}
\be H_{{tJV},j}^{(S)} = -t \left  [ {\bf S}_j {\bf S}_{j+1} +({\bf
S}_j {\bf S}_{j+1})^2\right ]+ {{2 V+3 t}\over 2} S_{z,j}^2
S_{z,j+1}^2 +{{J+ 2 t}\over 8} \left [ S_{+,j}^2 S_{-,j+1}^2 +
S_{-,j}^2 S_{+,j+1}^2+ 2 S_{z,j} S_{z,j+1} \right]
\label{H_tJV_S}\ee
\end{widetext}
from which the global Hamiltonian $\Hh_{tJV}^{(S)}$ is
straightforwardly obtained. \\The extended $t-J$ Hamiltonian in
the spin $1$ realization can be recognized as a sum of three
independent contributions. As a general comment, one may observe
that the interplay of such contributions to determine the ground
state phase diagram properties of $\Hh_{tJV}^{(S)}$ are easily
deduced from those of the corresponding fermionic model\cite{TTR}.
In particular, the $J$ term is expected to drive phase separation,
whereas the $V$ term is expected to be responsible of the opening
of a (spin) gapped phase. These phases are now to be interpreted
as driven from quadrupolar interaction in the spin description,
and in particular the gapped phase should be analyzed in terms of
some unusual quadrupolar ordering \cite{SOL}.

More specific interesting observations are now in order.\\
First of all --for arbitrary values of the three independent
parameters-- $\Hh_{tJV}^{(S)}$ inherits all the symmetries of its
fermionic partner. Since these were build in the even sector of
the on-site fermionic algebra, in order to give them we simply
have to translate ${\mathcal{N}}$ and $\vec{{\mathcal{S}}}$ in
terms of on-site projectors, and then to rewrite the latter into
their spin realization. Explicitly, they turn out to be
${\mathcal{N}} = \sum_jS_{z,j}^2$ , ${\mathcal{S_+}}= \sum_j
S_{+,j}^2$, ${\mathcal{S}}_-= {\mathcal{S}}_+^\dagger$, and
${\mathcal{S}}_z={1\over 2} \sum_j S_{z,j}$.\\Also, we notice that
the choice $J=-2 t$, $V=-{3\over 2}t$, which in the spin
realization correspond to the pure bilinear biquadratic spin $1$
Hamiltonian with $\Delta=1$ \cite{SOL}, in the fermionic
realization reads as a $t-J$ supersymmetric model extended by a
nearest neighbor repulsive term, in perfect agreement with
\cite{Ortiz}; in this case the absolute ground state of the latter
coincide with that given by Sutherland in \cite{SUT} for the
$SU(3)$ symmetric spin 1 model (also known as $F^3$ case (see
below)), and the spectrum is gapless.
\\Even more interestingly, there are other choices of parameters in
(\ref{tJV}) corresponding to integrable cases which generate
integrable spin 1 models not discussed in the literature. \\First
of all, the supersymmetric $t-J$ model in the spin realization
reads:\be H_{tJss}^{(S)}= -t\sum_j \left  [ {\bf S}_j {\bf
S}_{j+1} +({\bf S}_j {\bf S}_{j+1})^2 - 2  S_{z,j}^2 S_{z,j+1}^2
\right ]\quad ;\label{H_EBB}\ee this is the bilinear biquadratic
$\Delta=1$ spin Hamiltonian already cited, extended by a diagonal
nearest neighbor quadrupolar interaction. The ground state of the
latter is hence given by Lai-Sutherland\cite{SUT,LAI} solution of
supersymmetric $t-J$ model, and the spectrum is thoroughly
discussed in the literature.
\\Also surprising is the choice $J=V=0$, which in the fermionic
realization would correspond to the infinite $U$ Hubbard model:
its spectrum in one dimension is known \cite{ISCA} to be that of a
spinless fermion; in the spin $1$ realization we have that such is
also the spectrum of a non trivial model which to our knowledge
was never proved to be integrable.

\paragraph{Integrable spin $1$ models as generalized permutators}

Let's exploit more closely the above  observations. Thank to the
local character of our JWT, models proved to be integrable in one
language --which feature refers to the structure of the local
Hamiltonian-- are straightforwardly translated into integrable
models in the other language. In the following, we provide other
integrable spin $1$ models starting from fermionic ones.

It has recently been shown \cite{DOMO_NUCL} that both
supersymmetric $t-J$  and infinite $U$ electron models are
integrable in one dimension since they belong to a larger class of
models for which the local Hamiltonian's have the structure of
{\it generalized permutators}.\\
The meaning of a generalized permutator is easily understood in
terms of so called Sutherland species (SS). Starting from the
on-site vector space $V$, we may think to group its $d$ basis
vector (or {\it physical species}) into $N_S \leq d$ different
species which are called the Sutherland species. Each of these
species is left unchanged (apart from a possible sign change, see
below) by the action of the generalized permutator, the latter
interchanging only basis vector belonging to different SS's. A
generalized permutator would then have the structure of an
ordinary permutator if represented on a local vector space of
dimension $N_S$. \\Each of the $N_S$ SS is said bosonic ($B$) if
no sign change occurs after action of permutator, fermionic ($F$)
otherwise. For arbitrary $d$, there are many possibilities of
grouping the $d$ physical species into $N_S$ SS; each of them
corresponding to a different generalized permutator. The latter
can be classified as $B^{ l} F^{{k}-{l}}$ for $0\leq {l}\leq {k}$,
$1<{ k}\leq N_S$. Whenever the matrix representing the two-sites
Hamiltonian coincides with the matrix representation of a
generalized permutator, the global Hamiltonian is
integrable.\\
Returning to our case, for which $d=3$, it can be
seen\cite{DOMO_NUCL} that the infinite $U$ Hubbard model is a $BF$
model, where the bosonic species at each site is the vacuum, and
the fermionic one is formed by the two singly occupied states
(with up and down spin); whereas the supersymmetric $t-J$ model is
a $B F^2$ model, the three SS coinciding precisely with the
physical species at each site $|\alpha_F\rangle$. In general, it
has been shown that all integrable fermionic models with an
on-site vector space $V$ of dimension $3$, which locally act as
generalized permutator\cite{ADM}, and globally preserve
$\vec{\mathcal S}$ and ${\mathcal N}$ are eight; in correspondence
to the possible different choices $J=(s_1+s_2) t$, and $V=-{J\over
4}+(s_1+s_3) t$; with $s_\alpha=\pm 1$ for $\alpha=1,2,3$
independent signs. Our spin-fermion mapping allows now to map the
integrable fermionic cases into spin $1$ ones; interestingly,
apart from the cases with $J= \pm 2t$ already discussed in the
previous section, in so doing we obtain other $4$ integrable spin
$1$ models, which fact to our knowledge was never noticed. All of
these imply the choice $J=0$, thus corresponding to
generalizations of infinite $U$ Hubbard model. In the spin
realization their Hamiltonian reads
\begin{widetext}
\be H_{{EU\infty}}^{(S)} = - t \sum_j \left\{\left[ \left ( S_{+}
S_{z}\right )_j \left ( S_{z} S_{-}\right )_{j+1} + \left (S_{z}
S_{+}\right )_j \left (S_{-}S_{z}\right )_{j+1} +h.c.
\right]+(s_1+ s_2) S_{z,j}^2 S_{z,j+1}^2  \right\}\quad .
\label{H_EUinf_S} \ee
\end{widetext}
We can provide ground state energy $\epsilon=E_0/L$, with $E_0$
lowest eigenvalue, for each of these models.
\\In fact, following Sutherland notation, it is easily seen that
the models coincide --up to conserved quantities-- with a $B^2$
model for $s_1=s_2=-1$, with two $BF$ models for $s_1=-s_2$, and
with a $F^2$ model for $s_1=s_2=+1$. For all of them, $\epsilon$
can be evaluated using the extension of Sutherland theorem which
holds for generalized permutators \cite{DOMO_R}, and turns out to
be \bq \epsilon=\left\{
\begin{array}{lcc}
-1                                   & for & B^2 \\[5pt]
2 n_F -1 -{2\over \pi} \sin{\pi n_F} & for & BF  \\[5pt]
1-2\ln 2                             & for & F^2 \\
\end{array}
\right. \quad ; \eq here $n_F=N_F/L$ is the density of the
fermionic species, which is related to $\mathcal{N}$:
$N_F={\mathcal N}$ for $s_1=+1$, $N_F= L-{\mathcal N}$ for
$s_1=-1$.

\paragraph{Summary and conclusions.}
In this paper we revisited known spin-fermion mappings showing
that the underlying structure is that of local identities, which
relate different realizations of abstract projection operators.
This result is contained in equations (\ref{ham})-(\ref{op_prod}) and
holds for arbitrary dimension of both the ambient lattice
$\Lambda$ ($D$) and the on-site vector space $V$ ($d$), at the
very reasonable condition that the physical Hamiltonian is an even
operator in the fermionic fields. We then specialized to one
dimensional lattice, and in this case we explicitly gave our
generalized JWT in terms of simple local relations between
bilinear products of on-site projection operators in the spin and
fermionic languages (equation (\ref{JW_local})). The latter still holds
for arbitrary $d$, giving the standard JW results for $d=2$.
Finally we focussed on the case $d=3$, obtaining from the extended
fermionic $t-J$ models the corresponding $S=1$ models. In
particular, we used the mapping to generate new integrable spin
$1$ cases (equations (\ref{H_EBB}), (\ref{H_EUinf_S})) providing for
each of them the ground state energy.

Possible developments of present work are on the one hand in the
generalization of the local identity to other particle
realizations (for instance, hard-core bosons and anyons); as well
as in the explicit analysis of cases with $d>3$ (the
correspondence of extended Hubbard models with spin $3/2$ models
being of course the easier application). \\On the other hand in
providing a bridge to comprehension/solutions of models which
could strongly differ in their physical meaning, but are unified
from a mathematical point of view: for instance, even models in
dimension greater than one, like strips and ladders, can be mapped
into one-dimensional models with on-site vector space of
appropriate (finite) dimension avoiding the sign problem.\\
Work is in progress along these lines.

\end{document}